\documentclass[a4paper]{jpconf}
\usepackage{graphicx} 
\usepackage{amsmath}
\usepackage{hyperref}
 \usepackage{amsfonts}
 \usepackage{comment}
 \usepackage{float}

\usepackage{xcolor}

\begin{document}
\newcommand{\Nu}{\text{Nu}}
\newcommand{\Ra}{\text{Ra}}
\newcommand{\Par}{\text{Pr}}

\title{Hyperviscosity stabilisation of the RBF-FD solution to natural convection}

\author{Žiga Vaupotič$^1$, Miha Rot$^{1, 2}$ and Gregor Kosec$^1$}

\address{$^1$ Parallel and Distributed Systems Laboratory, Jožef Stefan Institute, Jamova Cesta 39, Ljubljana 1000, Slovenia}
\address{$^2$ Jožef Stefan International Postgraduate School, Jamova Cesta 39, Ljubljana 1000, Slovenia}

\ead{miha.rot@ijs.si}

\begin{abstract}
The numerical stability of fluid flow is an important topic in computational fluid dynamics as fluid flow simulations usually become numerically unstable in the turbulent regime. Many mesh-based methods have already established numerical dissipation procedures that dampen the effects of the unstable advection term. When it comes to meshless methods, the prominent stabilisation scheme is hyperviscosity. It introduces numerical dissipation in the form of a higher-order Laplacian operator. Many papers have already discussed the general effects of hyperviscosity and its parameters. However, hyperviscosity in flow problems has not yet been analyzed in depth. In this paper, we discuss the effects of hyperviscosity on natural convection flow problems as we approach the turbulent regime.
\end{abstract}

\section{Introduction}

Hyperviscosity was first introduced to spectral methods in 1998 by Ma under the name (super)spectral viscosity~\cite{ma1998superspectralviscosity}, however, the method itself has its roots in artificial viscosity, which was introduced by Von Neumman~\cite{VonNeumann1950}. Despite the promising analytical results shown by Ma, it was never widely used and only appeared in a handful of research papers where it was applied to the Navier-Stokes equation~\cite{chen2013,gelb2001}. The meshless community first adapted spectral viscosity, now called hyperviscosity, in 2011 when Fornberg~\cite{fornberg2011} showed that hyperviscosity was able to stabilise time-stepping schemes by shifting the spurious eigenvalues into the stable region. Subsequently, hyperviscosity was applied to a variety of fluid dynamic cases~\cite{barnett2015,flyer2016,flyer2012}. The parameters of hyperviscosity were initially only briefly mentioned and included a large number of empirical approximations. Hyperviscosity and its connection to the global stability of the Radial Basis Function-generated Finite Difference (RBF-FD) method was only well established by the work of Shankar~\cite{shankar2018, shankar2019}, who introduced the Von Neumann analysis and showed that hyperviscosity unconditionally stabilises linear advection-diffusion equations for multistep processes under the right parameters. Recently, hyperviscosity was also applied to nonlinear conservation laws and showed promising results~\cite{tominc2022}. In this paper, we address hyperviscosity in the context of a natural convection problem, i.e., a thermo-fluid problem, where we show that hyperviscosity can stabilise the solution of the standard De Vahl Davis benchmark problem solved on scattered nodes.  

\section{Numerical Method}
In order to numerically approximate the solution of the PDE on a domain $\Omega$ with the strong form RBF-FD~\cite{Tolstykh2003} method the domain is first discretised with a set of $x_i \in \Omega$ scattered nodes, using a dedicated meshless node positioning algorithm~\cite{Slak2019}. In next step the linear operator $\mathcal{L}$ applied to field $u$
\begin{align}
    (\mathcal{L} u) (x_c) \approx \sum_{i \in s_c} w_{i} u(x_i),
    \label{eq:linear_opr}
\end{align}
is approximated at the point $x_c$ using $n$ closest nodes, often referred to as stencil $s_c$.
The unknown RBF-FD weights $w$ are computed by imposing equality in Eq.~\eqref{eq:linear_opr} and solving the linear system for a set of radial basis functions. In our case that is a set of polyharmonic splines (PHS)~\cite{FlyerPoly2016} that are centred at the stencil nodes, augmented with polynomials up to
order $m$.
The PHS do not posses a shape parameter and are therefore seemingly without a free parameter. However, this is not completely true, as the selection of PHS order $k\in \mathbb{N}$ affects the stability of the system. In this work, we use the scaling $k= \ell + 1$ provided by Shankar~\cite{shankar2018}, where $\ell$ stands for the minimum unisolvency order i.e. the order of the highest derivative. The monomial order is set to the minimum $m = \ell$ as it is directly related to stencil size $n = 2 \binom{m+d}{m} + 1$, where $d=2$ is the dimensionality of the problem.
    
\section{Hyperviscosity}
To stabilise the higher frequency oscillations of the scheme we introduce the hyperviscosity term into the governing PDE. It helps by dampening spurious modes, usually stemming from the non-linear advection term, that would otherwise amplify with time.
The general form of the hyperviscosity term is 
\begin{align}
    (-1)^{1-\alpha}\gamma \Delta^\alpha u
    \label{eq:hyperviscosity}
\end{align}
where $\alpha$ is the order and $\gamma \ll 1$ is the  amplitude of applied hyperviscosity . The expression from Eq.~\eqref{eq:hyperviscosity} is added to the right-hand side of the equation at hand. The constant $\gamma$
has to be selected carefully to provide enough stabilisation without ruining the solution~\cite{flyer2012, fornberg2011}. There is no broad consensus in the literature with an abundance of proposed estimates and scalings for the constant $\gamma$~\cite{barnett2015, flyer2016, fornberg2011, shankar2018}.
We have decided to use a relatively simple scaling of $\gamma = c h^{2\alpha}$ \cite{flyer2012, flyer2016}, where $c$ is a user-defined constant. The paper by Flyer~\cite{flyer2016} suggests a bound of $\mathcal{O} (10^{-2})$ to $\mathcal{O} (10^1)$ for the constant $c$, however, we have noticed that the bound is case dependant prompting us to use higher values.

The order of hyperviscosity is also quite important, as higher orders of the operator tend to dampen higher frequency oscillations while having a less detrimental impact on the low-frequency behaviour i.e. the simulated physics~\cite{flyer2016}. Not to mention that the computational cost increases drastically with increasing order.  Based on the literature and the computational intensity of the hyperviscosity operator we use the order of $\alpha = 3$.  The order $\alpha = 3$ corresponds to $\nabla^6$ operator, while the rest of the problem is governed by  $\nabla^2$ and $\nabla$ operators, i.e. the order of approximation required for hyperviscosity is significantly higher in comparison to all other terms.  To improve computational performance and stability we therefore use separate approximation setups for the hyperviscosity term ( $k = 2\alpha + 1$ and $m = 2 \alpha$) and for other terms ($k = 3$ and $m = 2$).

\section{Governing problem}
We are solving a coupled system of buoyancy-driven incompressible Navier-Stokes and heat transfer equations,
\begin{subequations}
\label{eq:system}
    \begin{align}
        \label{eq:system1}
        \nabla \cdot u &= 0 \\
        \label{eq:system2}
        \frac{\partial u}{\partial t} + (u\cdot \nabla)u &= -\nabla p + \Pr\nabla^2 u - \Ra \Pr g \Delta T_r \\
        \label{eq:system3}
        \frac{\partial T}{\partial t} + (u \cdot \nabla) T &= \nabla^2 T,
    \end{align}
\end{subequations}
where $u$ stands for velocity and  $T$ for temperature , $g$ is the gravitational acceleration, $\Delta T_r$ is the temperature offset in Boussinesq approximation, $\Ra$ and $\Pr$ are nondimensional Rayleigh and Prandt numbers characterising the problem~\cite{DeVahlDavis1983}. 
The momentum equation is solved explicitly based on the temperature field of the previous iteration. The pressure-velocity coupling is implemented through the Chorin's pressure projection method~\cite{chorin1968}), where the pressure Poisson equation is solved implicitly, and the pressure correction is explicitly applied to the velocity field. The timestep of $\Delta t = 10^{-6}$ will be used throughout the whole analysis.

The hyperviscosity stabilisation term
\begin{equation}
    +c h^{6} \nabla^6 u,
\end{equation}
is added to the right hand side of the momentum Eq.~\eqref{eq:system2} and/or the heat Eq.~\eqref{eq:system3}.

\section{Results}
The results are based on the De Vahl Davis test case~\cite{DeVahlDavis1983}, where the natural convection is studied in a rectangular enclosed domain $\Omega = [0,1] \times [0,1]$ with no-slip velocity boundary conditions. The temperature gradient is imposed by Dirichlet boundary conditions for the temperature with $T_C = -0.5$ on the left wall and $T_H = 0.5$ on the right, while the horizontal walls are adiabatic. The Prandtl number is kept at $\Pr=0.71$ (air) throughout the whole analysis and the results are compared to 2 reference solutions~\cite{Kosec2011, Couturier2000} via the average Nusselt number along the cold wall $\partial \Omega_L$
\begin{align}
    \overline{\Nu} = \frac{1}{|\partial \Omega_L|}\sum_{x \in \partial \Omega_{L}} \frac{\partial T (x)}{\partial n}, 
\end{align}
used as the primary comparison indicator. Additionally, to analyze the stability of the system with the eigenvalue decomposition of the implicit (Euler) differentiation matrix $\mathbf{A} u^{n+1} = u^n$ is used.

\begin{figure}
\begin{center}
\includegraphics[width=\textwidth]{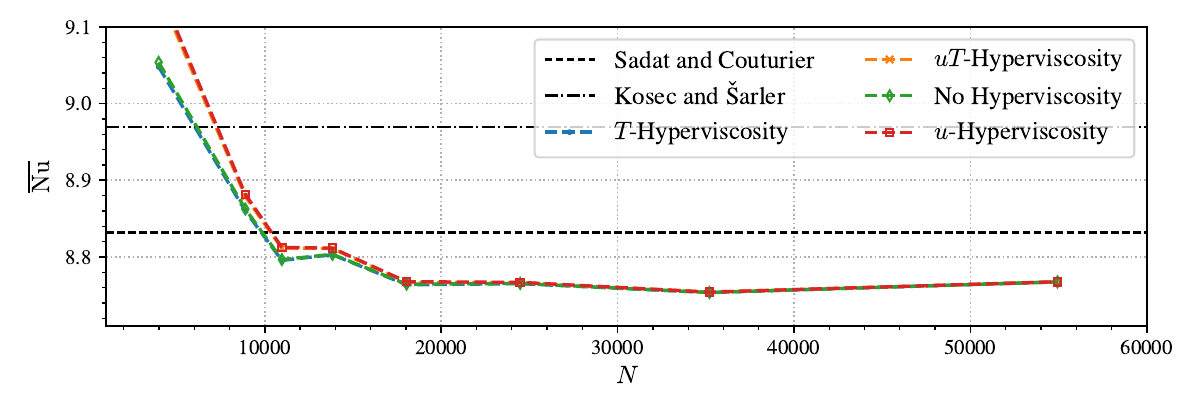}
 \end{center}
\caption{\label{nu-convergence}Convergence of the average Nusselt number on the left wall with an increasing number of computational nodes for $\Ra=10^6$. The line labelled as *-Hyperviscosity represents the results with additional hyperviscosity stabilisation for the specific equation. }
\end{figure}

We first verify our solution with the standard convergence test for $\Ra = 10^6$. The magnitude constant $c$ was hand-tuned to $c=100$, which will also be used in all following cases. In \autoref{nu-convergence} we see that the addition of hyperviscosity to the momentum equation, the temperature equation, or both equations, labelled as $u, T,$ and $uT$-Hyperviscosity, has almost no detrimental effect on the convergence. Our results are also in good agreement with the reference values indicated by the horizontal lines. The rest of the analysis will use a $\Delta x = 0.006$ ($N \approx 27 \cdot 10^4$) unless otherwise specified.

\begin{figure}
\begin{center}
\includegraphics[width=\textwidth]{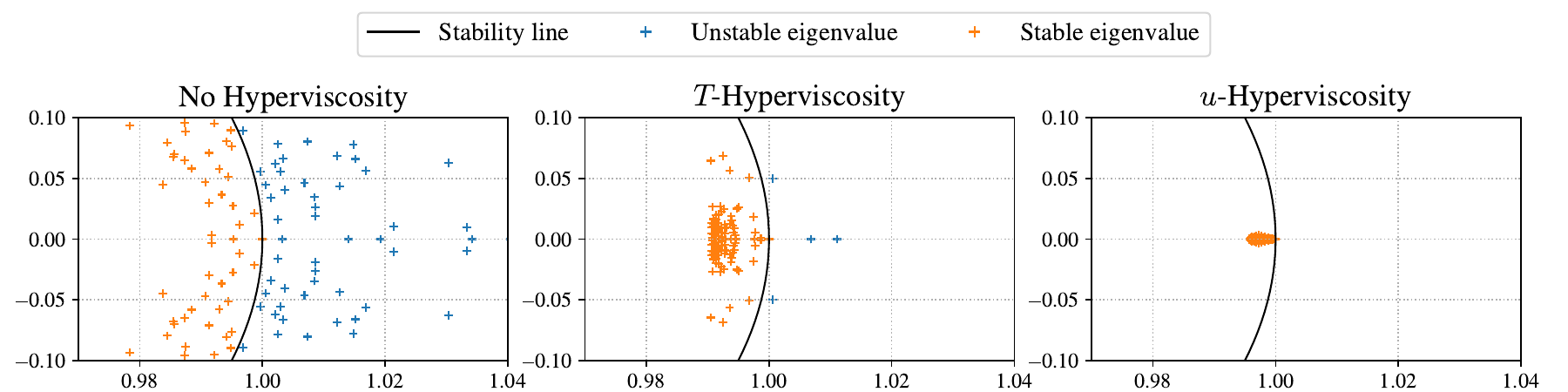}
\end{center}
\caption{\label{evaldecomp}Eigenvalue spectra of the inverse of implicit differentiation matrix $\mathbf{A}^{-1}$ at $t=3 \cdot 10^{-4}$ for $\Ra=10^8$ with spatial discretisation density of $\Delta x = 0.009$ ($N \approx 1.2 \cdot 10^4$). The black line shows a unit circle. }
\end{figure}

Up to this point, the results have shown that introducing hyperviscosity has little to no adverse effects on the accuracy of the numerical scheme, allowing us to continue the study of the impact on the stability of the system. 
First, we show that hyperviscosity stabilises the discrete dynamical system at $\Ra=10^8$. We find that in \autoref{evaldecomp} spectrum of the implicit differentiation matrix of the system, to which the hyperviscosity was applied, at $t=3 \cdot 10^{-4}$ has eigenvalues whose absolute values are less than one meaning that the linearized system is stable~\cite{Layek2015}, unlike the system without hyperviscosity. While the spectrum does not necessarily imply that the system is unconditionally stable, as the matrix $\mathbf{A}$ depends on the linearized advection term, it does, however, show that the system with hyperviscosity is stable for longer than the system without hyperviscosity. We also show that hyperviscosity must be applied to either the momentum equation or both the momentum and the heat transfer equations, as the stabilisation of the heat transfer equation alone doesn't shift all spurious eigenvalues to the stable region. 

\begin{figure}
\begin{center}
\includegraphics[width=\textwidth]{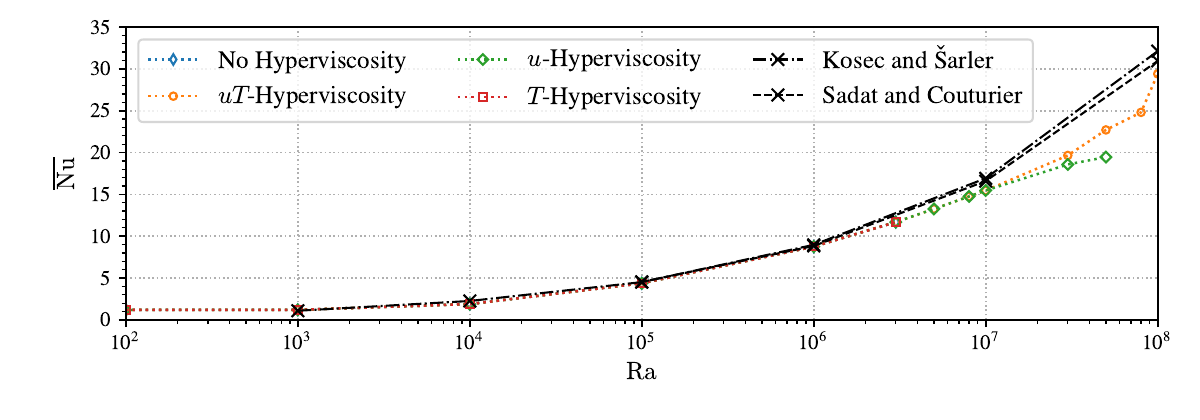}
\end{center}
\caption{\label{convera}The average left wall Nusselt number in the steady state for a range of Rayleigh numbers. The line labelled as *-Hyperviscosity represents the results with additional hyperviscosity stabilisation for the specific equation.}
\end{figure}

Furthermore, in \autoref{convera} we show that a system with hyperviscosity is far more stable for higher $\Ra$ numbers. We have also noticed that when the system approaches instability (as $\Ra \to 10^8$), the stabilisation of both the momentum and the heat transfer equations is required.
Interestingly, the stabilisation of the temperature field alone did not extend the stability to a higher $\Ra$ number. 

We have also tested a case where $Ra > 10^8$, but unfortunately no stable constant $c$ that could cover the entire considered range of cases was found. For optimal stabilisation effect, different values of the constant $c$ would be required for different $Ra$ and for different times in the solution procedure.

\begin{figure}[H]
\begin{center}
\includegraphics[width=\textwidth]{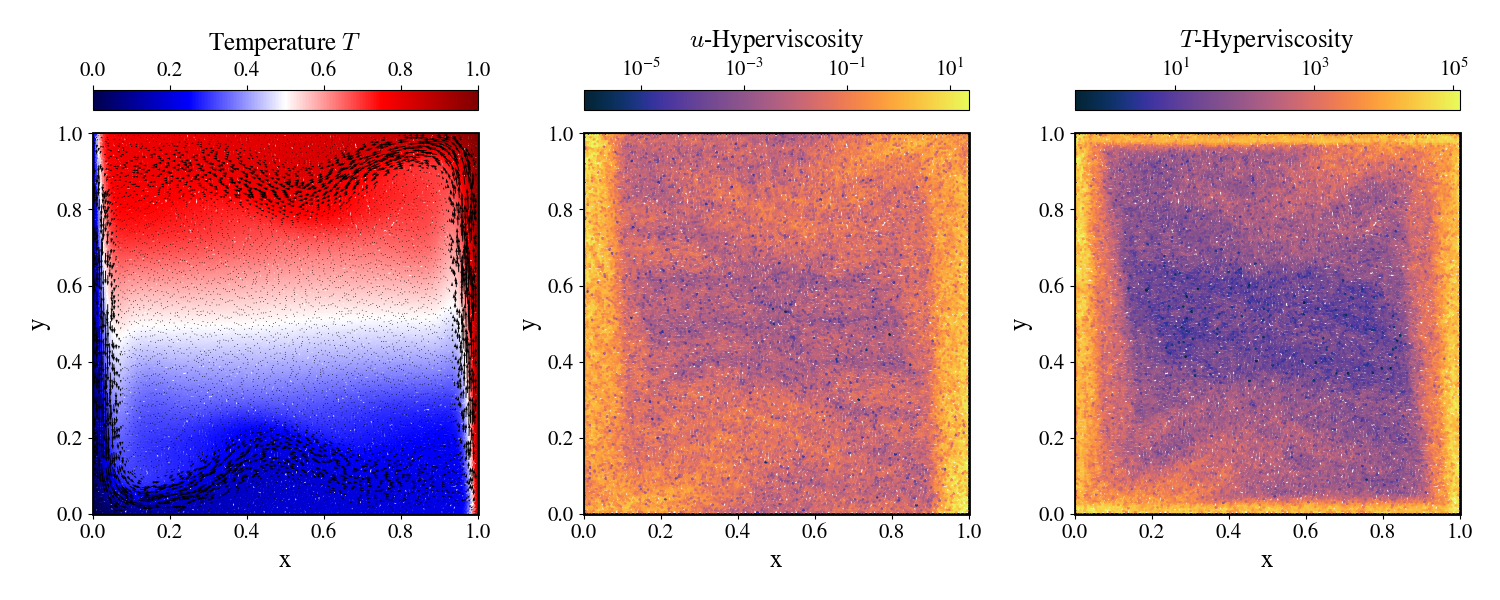}
\end{center}
\caption{\label{fields}Velocity and temperature field of the stationary solution \textit{(left)} at $Ra=10^7$ and $t=0.1$. Magnitudes of the hyperviscosity operator applied to the velocity field(\textit{center}) and the temperature field(\textit{right}).}
\end{figure}

As a final remark, an example of the resulting velocity and temperature fields is shown in the left subplot of \autoref{fields} and the corresponding stabilisation fields produced by applying the hyperviscosity operator to $u$ and $T$ in the central and right subplot. We can see that hyperviscosity is most effective in the areas with large velocity gradients for both and additionally close to the boundary for $T$-Hyperviscosity.

\section{Conclusions}
In this paper, we have shown that hyperviscosity stabilisation can be used to improve the stability for a natural convection-driven flow simulated with the meshless RBF-FD method on scattered computational nodes. We used a convergence study to establish that the introduction of hyperviscosity does not significantly change the solution once sufficiently dense discretisation is used.
Furthermore, we have shown how applying the stabilisation to momentum and heat equations changes the system's eigenspectrum and how these changes are reflected in the achievable range of Rayleigh numbers with stable solutions. By applying the hyperviscosity operator to both the momentum and the heat equation we were able to achieve comparable stability and results to reference meshless methods operating on regularly distributed computational nodes.

The main challenge encountered has proved to be the selection of hyperviscosity parameters, most critically the constant $\gamma$ and its scaling relations that ensure adequate stabilisation in the wide considered range of flow regimes and discretisation densities. This is also the main hurdle for further practical applications of hyperviscosity stabilisation where a method for adapting $\gamma$ to specific cases and even to changing flow conditions within a single case would be sorely required.
%\vspace{-4mm}
\ack
The authors acknowledge the financial support from the Slovenian Research and Innovation Agency (ARIS) research core funding No. P2-0095, Young Researcher programme PR-10468, and research projects No. J2-3048 and No. N2-0275.

Funded by National Science Centre, Poland under the OPUS call in the Weave programme 2021/43/I/ST3/00228.
This research was funded in whole or in part by National Science Centre (2021/43/I/ST3/00228). For the purpose of Open Access,
the author has applied a CC-BY public copyright licence to any Author Accepted Manuscript (AAM) version arising from this submission.
%\vspace{-4mm}
\section*{References}
\bibliography{lit}
\bibliographystyle{iopart-num}
\end{document}